\documentclass[a4paper]{article}
\usepackage{nnsp2e}
\usepackage{fancybox,color}
\usepackage{xspace}
\usepackage{epic,eepic}
\usepackage{bbm}
\usepackage{amssymb}            
\usepackage{graphicx}            
\usepackage[T1]{fontenc}        
\input{alphabet2.tex}
\input{abrege2.tex}
\def\bigcro#1{\bigl[#1\bigr]}

                \def\Esp#1{{\mathrm{E}}\bigcro{#1}}

\newsavebox{\fminibox}
\newlength{\fminilength}
\newenvironment{fminipage}[1][\linewidth]
  {\setlength{\fminilength}{#1}
   \begin{lrbox}{\fminibox}\begin{minipage}{\fminilength}}
  {\end{minipage}\end{lrbox}\noindent\fbox{\usebox{\fminibox}}}

  \def\+{^\dagger}

\def\nequiv{\not\kern-.05em\equiv}
\def\egal{\kern-.5em=\kern-.5em}        
\def\propt{\kern-.2em\propto\kern-.2em} 

\def\intdouble{\int\kern-0.3em\int}
\def\inttriple{\int\kern-0.3em\int\kern-0.3em\int}

\def\rond#1{\overset{\kern-0.33em~_\circ}{#1}}
\def\rondit[#1]#2{\overset{\kern#1~_\circ}{#2}}

\def\babs{\begin{abstract}}             \def\eabs{\end{abstract}}
\def\barr{\begin{array}}                \def\earr{\end{array}}
\def\bcc{\begin{center}}                \def\ecc{\end{center}}

\def\bdes{\begin{description}}          \def\edes{\end{description}}
\def\bdoc{
\begin{document}}             \def\edoc{\end{document}}
\def\ben{\begin{enumerate}}             \def\een{\end{enumerate}}
\def\beqn{\begin{eqnarray}}             \def\eeqn{\end{eqnarray}}
\def\beqnl#1{\beqn\label{#1}}           \def\eeqnl#1{\label{#1}\eeqn}
\def\beqnx{\begin{eqnarray*}}           \def\eeqnx{\end{eqnarray*}}
\def\bseqn{\begin{subeqnarray}}         \def\eseqn{\end{subeqnarray}}
\def\beq#1\eeq{\begin{equation}#1\end{equation}}
\def\bal#1\eal{\begin{align}#1\end{align}}
\def\balx#1\ealx{\begin{align*}#1\end{align*}}
\def\beqx{$$}                           \def\eeqx{$$}
\def\bfig{\protect\begin{figure}}       \def\efig{\protect\end{figure}}
\def\bfigx{\protect\begin{figure*}}     \def\efigx{\protect\end{figure*}}
\def\bfigt{\protect\begin{figurette}}   \def\efigt{\protect\end{figurette}}
\def\bfl{\begin{flushleft}}             \def\efl{\end{flushleft}}
\def\bfr{\begin{flushright}}            \def\efr{\end{flushright}}
\def\bit{\begin{itemize}}               \def\eit{\end{itemize}}
\def\bmi{\begin{minipage}}              \def\emi{\end{minipage}}
\def\bfmi{\begin{fminipage}}            \def\efmi{\end{fminipage}}
\def\bpic{\begin{picture}}              \def\epic{\end{picture}}
\def\bqu{\begin{quote}}                 \def\equ{\end{quote}}
\def\bqun{\begin{quotation}}            \def\equn{\end{quotation}}
\def\bsl{\begin{slide}}                 \def\esl{\end{slide}}
\def\btabb{\begin{tabbing}}             \def\etabb{\end{tabbing}}
\def\btabl{\begin{table}}               \def\etabl{\end{table}}
\def\btablx{\begin{table*}}             \def\etablx{\end{table*}}
\def\btab{\begin{tabular}} 
\def\btabu{\begin{tabular}}             \def\etabu{\end{tabular}}
\def\btabx{\begin{tabular*}}            \def\etabx{\end{tabular*}}
\def\bbib{\begin{thebibliography}{999}} \def\ebib{\end{thebibliography}}
\def\bver{\begin{verbatim}}             \def\ever{\end{verbatim}}
\def\bca{\begin{cases}}                          \def\eca{\end{cases}}

\def\Rbe{\Rb_{\epsilon}}
\def\seps{\sigma_{\epsilon}}
\def\seta{\sigma_{\eta}}
\def\Gz{\Gammab_{z}}   
\title{MCMC joint separation and segmentation of hidden Markov fields}
\author{
\HS and \AMD\\
\LSS (\tutelle). \\
\adresse. \\
E-mail: \texttt{snoussi@lss.supelec.fr, \, djafari@lss.supelec.fr}}


\bdoc
\maketitle

\begin{abstract}
In this contribution, we consider the problem of the  blind separation of noisy instantaneously mixed  images. The images are modelized by hidden Markov fields with unknown parameters. Given the observed images, we give a Bayesian formulation and we propose to solve the resulting data augmentation problem  by implementing a Monte Carlo Markov Chaîn (MCMC) procedure. We separate the unknown variables into two categories: \\
$1$. The parameters of interest which are the mixing matrix, the noise covariance and the parameters of the sources distributions.\\
$2$. The  hidden variables which are the  unobserved sources and the unobserved pixels classification labels.

The proposed algorithm provides in  the stationary regime samples drawn from the posterior distributions of all the variables involved in the problem leading to a flexibility in the cost function choice.

We discuss and characterize  some problems of non identifiability and degeneracies of the parameters likelihood and the behavior of the MCMC algorithm in this case. 

Finally, we show the results for both synthetic and real data  to illustrate the feasibility of the proposed solution.
\end{abstract}
%
%
\section{I. Introduction and model assumptions}
The observations are $m$ images $(\Xb^{i})_{i=1..m}$, each image $\Xb^{i}$ is defined on a finite set of sites, $\Sc$, corresponding to the pixels of the image: $\Xb^{i}=(x_{r}^{i})_{r \in \Sc}$. The observations are noisy linear instantaneous mixture of $n$ source images $(\Sb^{j})_{j=1..n}$ defined on the same set $\Sc$:
\[
x_{r}^{i}=\sum_{j=1}^{n} a_{ij} s_{r}^{j} + n_{r}^{i}, \; r \in \Sc, i=1..m
\]
where $\Ab=(a_{ij})$ is the unknown mixing matrix, $\Nb^{i}=(n_{r}^{i})_{r \in \Sc}$ is a zero-mean white Gaussian noise with variance ${\seps}_{i}^{2}$. 
At each site $r \in \Sc$, the matrix notation is:
\beq\label{bss}
\xb=\Ab\,\sb + \nb
\eeq
The noise and source components $(\Nb^{i})_{1..m}$  and $(\Sb^{j})_{j=1..n}$  are supposed to be independent. Each source is modelized by a double stochastic process $(\Sb^j, \Zb^j)$. $\Sb^j$ is a field of  values in a  continuous set $\Rc$ and represents the real observed image in the absence of noise and mixing deformation. $\Zb^{j}$ is the hidden Markov field representing the unobserved pixels classification whose components are in a discrete set, $\Zb_{r}^{j} \in \{1..K^j\}$. The joint probability distribution of $\Zb^{j}$ satisfies the following properties,
\[\left\{\barr{ll}
\forall \, \Zb^{j}, & P_{M}(z_{r}^{j} \,|\, \Zb_{\Sc \backslash \{r\}}^{j})= P_{M}(z_{r}^{j} \,|\, \Zb_{N(r)}^{j}) \\
~\\
\forall \, \Zb^{j}, & P_{M}(\Zb^{j}) > 0
\earr \right.
\]
where $\Zb_{\Sc \backslash \{r\}}^{j}$ denotes the field restricted to $\Sc \backslash \{r\}=\{l \in \Sc, l \neq r\}$ and $N(r)$ denotes the set of neighbors of $r$, according to the neighborhood system defined on  $\Sc$ for each source component. According to the Hammersley-Clifford theorem, there is an equivalence between a Markov random field and a Gibbs distribution,
\[
P_{M}(\Zb^{j})=[W(\alpha_j)]^{-1} \exp\{-H_{\alpha_j}(\Zb^{j})\}
\]
where $H_{\alpha_j}$ is the energy function and $\alpha_j$ is a parameter weighting the spatial dependencies supposed to be known. Conditionally to the hidden discrete field $\Zb^{j}$, the source pixels $S_{r}^{j}, r \in \Sc$ are supposed to be independent and have the following conditional distribution:
\[
p(\Sb^{j} \,|\, \Zb^{j}, \etab^{j}) = \prod_{r \in \Sc} p_{r}(s_{r}^{j} \,|\, z_{r}^{j}, \etab^{j})
\]
where the positive conditional distributions depend on the parameter $\etab^{j} \in \Rc^{d}$. We assume in the following that  $p_{r}(. \,|\, z)$ is a Gaussian distribution with parameters $\etab^{j}=(\mu_{jz}, \sigma_{jz}^2)_{z=1..K}$.

We note that we have a two-level inversion problem: 
\begin{enumerate}
\item The problem described by (\ref{bss}) when the mixing matrix $\Ab$ is unkown is the source separation problem \cite{Cardoso97,Knuth99, Djafari99d}. 
\item Given the source component $\Sb^{j}$, the estimation of the parameter $\etab^{j}$ and the recovering of  the hidden classification $\Zb^{j}$  is known as the unsupervised segmentation \cite{Peyrard01}. 
\end{enumerate}
In this contribution, given the observations $\Xb^{i}, i=1..m$, we propose a solution to jointly separate the $n$ unknown sources and perform their  unsupervised segmentations. In section II, we give a Bayesian formulation of the problem. In section III, an MCMC algorithm based on the data augmentation modelization is proposed. In section IV, we focus on the problem of the non identifiability and the degeneracies occurring in the source separation problem and their effects on the MCMC implementation. In section V, numerical simulations are shown to illustrate the feasibility of the solution.   

\section{II. Bayesian formulation}
Given the observed data $\Xb=(\Xb^1,...,\Xb^m)$, our objective is the estimation of the mixing matrix $\Ab$, the noise covariance $\Rbe=diag({\seps}_{1}^2,...,{\seps}_{m}^2)$, the means and variances $(\mu_{jz}, \sigma_{jz}^2)_{j=1..n, z=1..K}$ of the conditional Gaussians of the prior distribution of the sources. The \apost distribution of the whole parameter $\thetab=(\Ab, \Rbe, \mu_{jz}, \sigma_{jz}^2)$ contains all the information that we can extract from the data. According to the Bayesian rule, we have
\[
p(\thetab \,|\, \Xb)  \propto p(\Xb \,|\, \thetab) p(\thetab)
\]
In the section III, we will discuss the attribution of appropriate prior distribution $p(\thetab)$. Concerning the likelihood, it has the following expression,
\beq\barr{lll}\label{post}
p(\Xb \,|\, \thetab)&=&\displaystyle{\sum_{\Zb}\int_{\Sb} p(\Xb, \Sb, \Zb \,|\, \thetab) d\,\Sb}\\
~\\
~&=& \displaystyle{\sum_{\Zb}   \left\{ \prod_{r \in \Sc} \Nc(\xb_{r}\,;\, \Ab\mub_{\zb_r}, \Ab\Rb_{\zb_r}\Ab^{*}+\Rbe) \right\} P_{M}(Z)}
\earr
\eeq
where $\Nc$ denotes the Gaussian distribution, $\xb_r$ the $(m \times 1)$ vector of observations on the site $r$, $\zb_r$ is the vector label, $\mub_{\zb_r}=[\mu_{1z_1},...,\mu_{nz_n}]^{t}$ and  $\Rb_{\zb_r}$ the diagonal matrix $diag[\sigma_{1z_1}^2,...,\sigma_{nz_n}^2]$. We note that the expression (\ref{post}) hasn't  a tractable form with respect to the parameter $\thetab$ because of the integration of the hidden variables $\Sb$ and $\Zb$. This remark leads us to consider the data augmentation algorithm \cite{Tanner87} where we complete the observations $\Xb$ by the hidden variables $(\Zb, \Sb)$, the complete data are then $(\Xb, \Sb, \Zb)$. In a previous work \cite{Snoussi01d}, we implemented restoration maximization algorithms in the one dimensional case to estimate the maximum \apost estimate of $\thetab$. We extend this work in two directions: {\bf (i)} first, the sources are two-dimensional signals, {\bf (ii)} second, we implement an MCMC algorithm to obtain samples of $\thetab$ drawn from its \apost distribution. This gives the possibility of not being restricted to estimate the parameter by its maximum \apost, we can consider another cost function and compute the corresponding estimate.

\section{III. MCMC implementation}
We divide the vector of unknown variables into two sub-vectors: The hidden variables $(\Zb, \Sb)$  and the parameter $\thetab$ and we consider a Gibbs sampler:

\vspace{0.5cm}
repeat until convergence,
\begin{enumerate}
\item draw $(\tilde{\Zb}^{(k)}, \tilde{\Sb}^{(k)}) \thicksim p(\Zb, \Sb \,|\, \Xb, \tilde{\thetab}^{(k-1)})$
\item draw $\tilde{\thetab}^{(k)} \thicksim  p(\thetab \,|\, \Xb, \tilde{\Zb}^{(k)}, \tilde{\Sb}^{(k)})$
\end{enumerate}

This Bayesian sampling \cite{Robert96} produces a Markov chain $(\tilde{\thetab}^{(k)})$,  ergodic with stationary distribution $p(\thetab \,|\, \Xb)$. After $k_0$ iterations (warming up), the samples $(\tilde{\thetab}^{(k_0+h)})$ can be considered to be drawn approximately from their \apost distribution $p(\thetab \,|\, \Xb)$. Then, by the ergodic theorem, we can approximate \apost expectations by empirical expectations:
\beq\label{emp}
\Esp{h(\thetab) \,|\, \Xb} \approx \frac{1}{K} \sum_{k=1}^{K} h(\tilde{\thetab}^{(k)})
\eeq 

\noindent {\bf Sampling $(\Zb, \Sb)$:}
The  sampling of  the hidden fields $(\Zb, \Sb)$ from $p(\Zb, \Sb \,|\, \Xb, \thetab)$ is obtained by,
\begin{enumerate}
\item draw $\tilde{\Zb}$ from 
\[
p(\Zb \,|\, \Xb, \thetab) \propto p(\Xb \,|\, \Zb, \thetab)\,P_{M}(\Zb)
\] 
In this expression, we have two kinds of dependencies: $\bf (i)$ $\Zb$ are independent across components, $p(\Zb)=\prod_{j=1}^{n} p(\Zb^j)$ but each discrete image $\Zb^j \thicksim P_{M}(\Zb^j)$ has a Markovian structure. $\bf (ii)$ Given $\Zb$, the fields $\Xb$ are independent through the set $\Sc$, $p(\Xb \,|\, \Zb, \thetab)=\prod_{r \in \Sc} p(\xb_r \,|\, \zb_r, \thetab)$ but dependent through the components because of the mixing operation $p(\xb_r \,|\, \zb_r, \thetab)=\Nc(\xb_{r}\,;\, \Ab\mub_{\zb_r}, \Ab\Rb_{\zb_r}\Ab^{*}+\Rbe)$.

\item draw $\tilde{\Sb} \,|\, \tilde{\Zb}$ from
\[
p(\Sb \,|\, \Xb, \Zb, \thetab) = \prod_{r \in \Sc} \Nc(\sb_r \,;\,\mb_r^{apost}, \Vb_r^{apost} )
\]
where the \apost mean and covariance are easily computed \cite{Snoussi00a},
\[\barr{lll}
\Vb_r^{apost} &=& \left[ \Ab^{*}\Rbe^{-1}\Ab + \Rb_{\zb_r}^{-1} \right]^{-1} \\
~\\
\mb_r^{apost} &=& \Vb_r^{apost} \left( \Ab^{*} \Rbe^{-1} \xb_r + \Rb_{\zb_r}^{-1} \mub_{\zb_r} \right) 
\earr
\]
\end{enumerate}

\noindent {\bf Sampling $\thetab$:}
Given the observations $\Xb$ and the samples $(\Zb, \Sb)$, the sampling of the parameter $\thetab$ becomes an easy task (this represents the principal reason of introducing the hidden sources). The conditional distribution $p(\thetab \,|\, \Xb, \Zb, \Sb)$ is factorized into two conditional distributions,
\[
p(\thetab \,|\, \Xb, \Zb, \Sb) \propto p(\Ab, \Rbe \,|\, \Xb, \Sb) \, p(\mub, \sigmab  \,|\, \Sb, \Zb)
 \] 
 leading to a separate sampling of $(\Ab, \Rbe)$ and $(\mub, \sigmab)$. The choice of the \aprio distributions is not an easy task \cite{Kass94}. The complete likelihood of $(\Ab, \Rbe)$ belongs to the location scale family \cite{Box72} and applying the Jeffrey's rule we have,
 \[
 p(\Ab, \Rbe) \propto | \Fc(\Rbe) |^{\frac{1}{2}}=|\Rbe|^{\frac{-(m+1)}{2}}
 \]
 where $p(\Ab)$ is locally uniform and $\Fc$ is the Fisher information matrix. We obtain an inverse Wishart distribution for the noise covariance and a Gaussian distribution for the mixing matrix,
 \[\left\{ \barr{l}
 \Rbe^{-1} \thicksim Wi_{m}(\alpha_{\epsilon}, \betab_{\epsilon}), \; \alpha_{\epsilon}=\frac{|\Sc|-n}{2}, \;\betab_{\epsilon}= \frac{|\Sc|}{2}(\Rb_{xx}-\Rb_{xs}\Rb_{ss}^{-1}\Rb_{xs}^{*}) \\
 ~\\
 p(\Ab \,|\, \Rbe) \thicksim \Nc(\mub_a, \Rb_a), \mub_a=vec(\Rb_{xs} \Rb_{ss}^{-1}), \; \Rb_a=\frac{1}{|\Sc|} \Rb_{ss}^{-1} \otimes \Rbe
 \earr \right.
 \]   
 where we define the empirical statistics $\Rb_{xx}=\frac{1}{|\Sc|}\sum_{r} \xb_r \xb_r^{*}$, $\Rb_{xs}=\frac{1}{|\Sc|}\sum_{r} \xb_r \sb_r^{*}$ and $\Rb_{ss}=\frac{1}{|\Sc|}\sum_{r} \sb_r \sb_r^{*}$. We note that the covariance matrix of $\Ab$ is proportional to the noise to signal ratio. This explains the fact noted in \cite{Bermond00} concerning the slow convergence of the EM algorithm. 
For the parameters $(\mub, \sigmab)$, we choose conjugate priors \cite{Robert96}. The reason of this choice is the elimination of degeneracies occurring when estimating the variances  $ \sigma_{jz}$. This point is elucidated in section IV. The \apost distribution remains in the same family as the likelihood function, Gaussian for the means and Inverse Gamma for the variances. The expressions are the same as in \cite{Robert96}. 

\section{IV. Identifiability and degeneracies}
It is well known that in the source separation problem  there exist scale and permutation indeterminations. This can be seen when multiplying the matrix $\Ab$ by a scale permutation matrix $\Lambda \Pb$ and the sources by $\Pb^{T} \Lambda^{-1}$. The permutation indetermination doesn't degrade the performance of the algorithm. In fact, in image processing the size of data $|\Sc|$ is sufficiently large to  avoid the Markov chaîn $(\tilde{\Ab}^{(k)})$  produced by the algorithm permuting its columns,  the probability that the Markov chain changes the \apost mode is very low. However, the scale indetermination must be eliminated. In practice, after each iteration of the MCMC algorithm, the columns of $\Ab$ are normalized. 

There is another kind  of indetermination  which is the transfer of variances between the covariances $\Rb_z$ and the noise covariance $\Rbe$:
\[
p(\Xb \,|\, \Ab, \Rbe+\epsilon \Ab \Lambda \Ab^{*}, \Rb_z-\epsilon \Lambda, \mub_z)=p(\Xb \,|\, \Ab, \Rbe, \Rb_z, \mub_z), \; \epsilon=\pm1
\]   
When we study the particular case of   diagonal noise covariance and the   mixing matrix $\Ab$ is unitary  ($\Ab \,\Ab^{*}=\Ib$), we note that an obvious  transfer of variances occurs when $\Lambda = \alpha \Ib$.   A retained  solution in this paper is the penalization of the likelihood by a prior on the variances which eliminates this variance transfer. This solution is more robust than the fact of fixing either $\Rb_z$ or $\Rbe$. However, a simultaneous penalization of noise and signal variance can induce a transfer between modes. In such situations, the Markov  chains $(\tilde{\Rbe}^{(k)})$ and $(\tilde{\Rb_z}^{(k)})$ seem to converge to a stationary distribution even after a great number of iterations but suddenly a transfer occurs (see the section V for numerical illustration). This indetermination is noted in \cite{Bermond00} and was used to accelerate the convergence of the EM algorithm by forcing the noise covariance to be maximized. In the MCMC algorithm, we note that the \apost covariance of $\Ab$ is $\Rb_a=\frac{1}{|\Sc|} \Rb_{ss}^{-1} \otimes \Rbe$. Consequently, as the signal to noise  ratio increases, covariance decreases and the sample $\tilde{\Ab}$ is more concentrated on its mean value.   

It is obvious, under the form \ref{post}, that degeneracy happens when one of the terms constituting the sum approaches to infinity  and this is  independent of the  law  $P_{M}$.

Consider now the matrices $\Gz=\Ab\Rb_z\Ab^{*}+\Rbe$. It's clear that degeneracy is produced when, among matrices $\Gz$, at least one is singular  and  one is regular. We  show in the following that this situation can  occur.

We recall that the matrices $\Rb_z$ and $\Rbe$ belong to a closed subset of the set of the non negative definite  matrices. Constraining matrices to be positive definite leads to complicated solutions. The main origin  of this complication is the fact that the set of positive definite matrices is not closed. For the same reason, we don't constrain the mixing matrix $\Ab$ to be of full rank.

\vspace{0.3cm}
{\bf Proposition $1$}: $\forall$ $\Ab$ non null, $\exists$ matrices $\{ \Gz=\Ab\Rb_z\Ab^{*}+\Rbe$ for  $z=1..K$\} such that $\{z \;|\; \Gz \; is \; singular\} \neq \emptyset$ and $\{z \;|\; \Gz \; is \; regular\} \neq \emptyset$. \\
$\Rbe$ is necessarily a singular NND matrix and $Card\left(\{z \;|\; \Rb_z \; is \; regular\}\right) < K$.
\vspace{0.3cm}
 
For a detailed proof see \cite{Snoussi01c}

\vspace{0.1cm}
One possible way to eliminate this degeneracy  consists in penalizing the likelihood by an  Inverse Wishart \aprio for covariance matrices. In fact, we know that the origin of degeneracy is that the covariance matrices  $\Rb_z$ and $\Rbe$ approach the boundary of singularity (in a non arbitrary way). Thus, if we  penalize the likelihood such that  when one of the covariance matrices approaches the boundary, the \apost distribution goes to zero, eliminating  the infinity value at the boundary and even forcing it to zero.

\vspace{0.3cm}
{\bf Proposition $2$:} $\forall$ $\Xb$ $\in$ $(\RD^{m})^{|\Sc|}$, the likelihood  $p(\Xb\,|\,\thetab)$ penalized by an \aprio Inverse Wishart for the noise covariance matrix $\Rbe$ or by  an \aprio Inverse Wishart for the matrices $\Rb_z$ is bounded and goes to  $0$ when one of the covariance matrices approaches the boundary of singularity.
\vspace*{0.3cm} 

For a detailed proof see \cite{Snoussi01c}. 

\section{V. Simulation results}
To illustrate the feasibility of the algorithm, we generate two discrete fields of $64 \times 64$ pixels from the Ising model,
\[
P_M(\Zb^{j})=\left[ W(\alpha_j) \right]^{-1} \exp\{\alpha_j \sum_{r \thicksim s}I_{z_r=z_s}\}, \alpha_1=2, \; \alpha_2=0.8
\] 
$\alpha_1 > \alpha_2$ implies that the first source is more homogeneous than the second source. Conditionally to $\Zb$, the continuous sources  are generated from Gaussian distributions of means  $\mu_{jz}=\left[\barr{ll} -2 & 2 \\ -3 & 3 \earr \right]$ and variances $\sigma_{jz}=\left[\barr{ll} 1 & 2 \\1 & 2\earr \right]$. 


The sources are then mixed with the matrix $\Ab=\left[\barr{ll} 0.85 & 0.44 \\0.51 & 0.89\earr \right]$ and a white Gaussian noise with covariance $\seps^2 \Ib$ ($\seps^2=5$) is added. The signal to noise ratio is $1$ to $3$ dB.  Figure-$1$ shows the mixed signals obtained on the detectors. 

\setlength{\tabcolsep}{0.4cm}
\btabu{l}
\includegraphics[width=100mm,height=50mm]{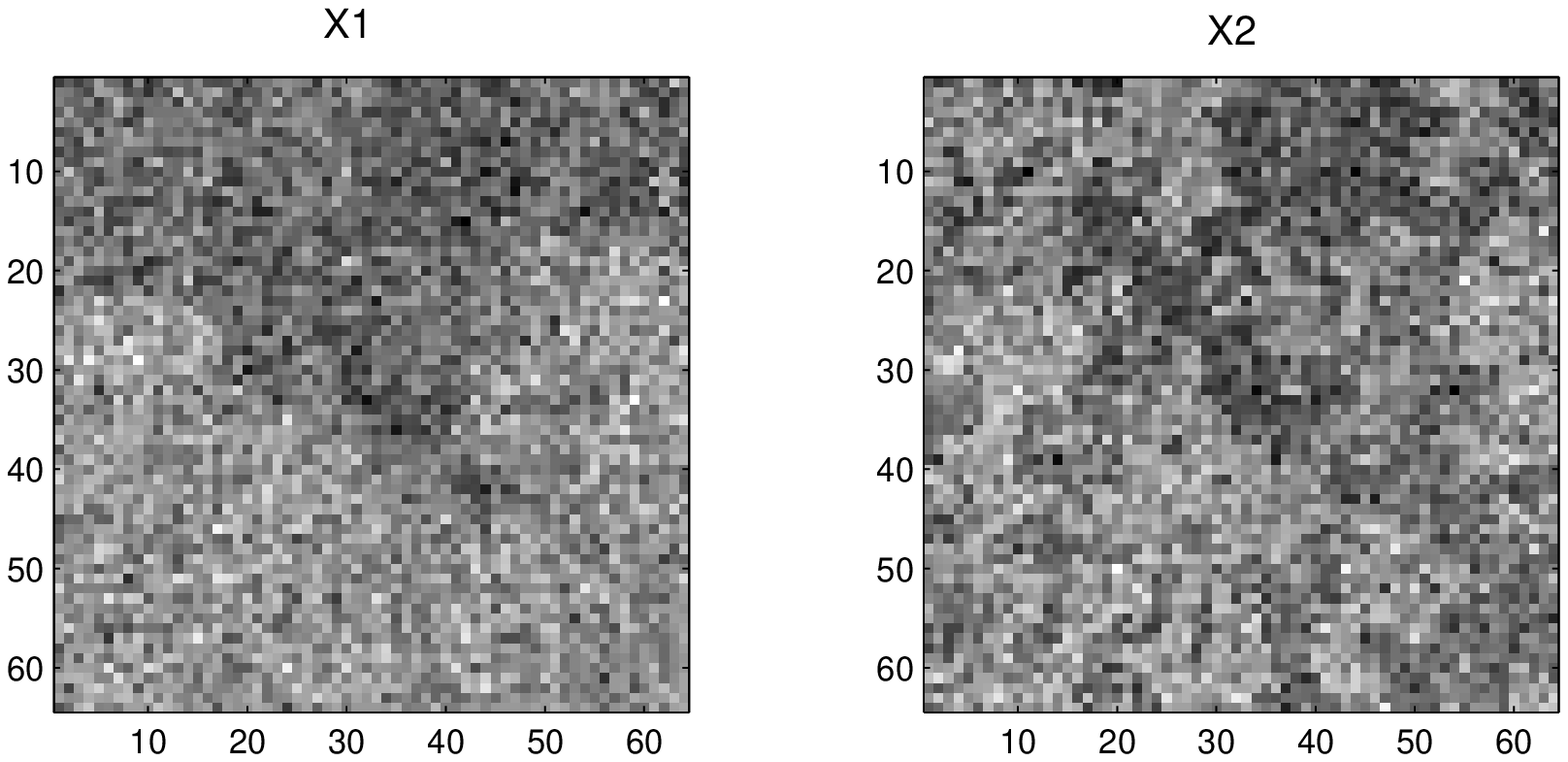} \\
\hspace{1.2cm} {\small figure-$1$. The noisy mixed  images $\Xb^1$ and $\Xb^2$}
\etabu

\vspace{1cm}
We apply the MCMC algorithm described in section III to obtain the Markov chaîns $\Ab^{(k)}$, $\Rbe^{(k)}$, $\mu_{jz}^{(k)}$ and ${\sigma_{jz}^{2}}^{(k)}$. Figures $2$ and $3$ show the histograms of the element samples of $\Ab$ and their empirical expectations (\ref{emp}). We note the concentration of the histograms representing approximately the marginal distributions around the true values and the convergence of the empirical expectations after about $1000$ iterations. Figures $4$ and $5$ show the convergence of the empirical expectations. We note that the convergence of the variances is slower that the mixing elements and the means.  Figure $6$ shows the transfer of variances when the matrix $\Ab\left[\barr{ll} 0.89 & -0.44 \\0.44 & 0.89\earr \right]$ is unitary. We note that this transfer occurred after a great number of iterations ($80000$ iterations) and that the sum of the variances remains constant.

\setlength{\tabcolsep}{0.4cm}
\btabu{@{\hspace{0.1mm}}c@{\hspace{0.1mm}}c}
\includegraphics[width=70mm,height=55mm]{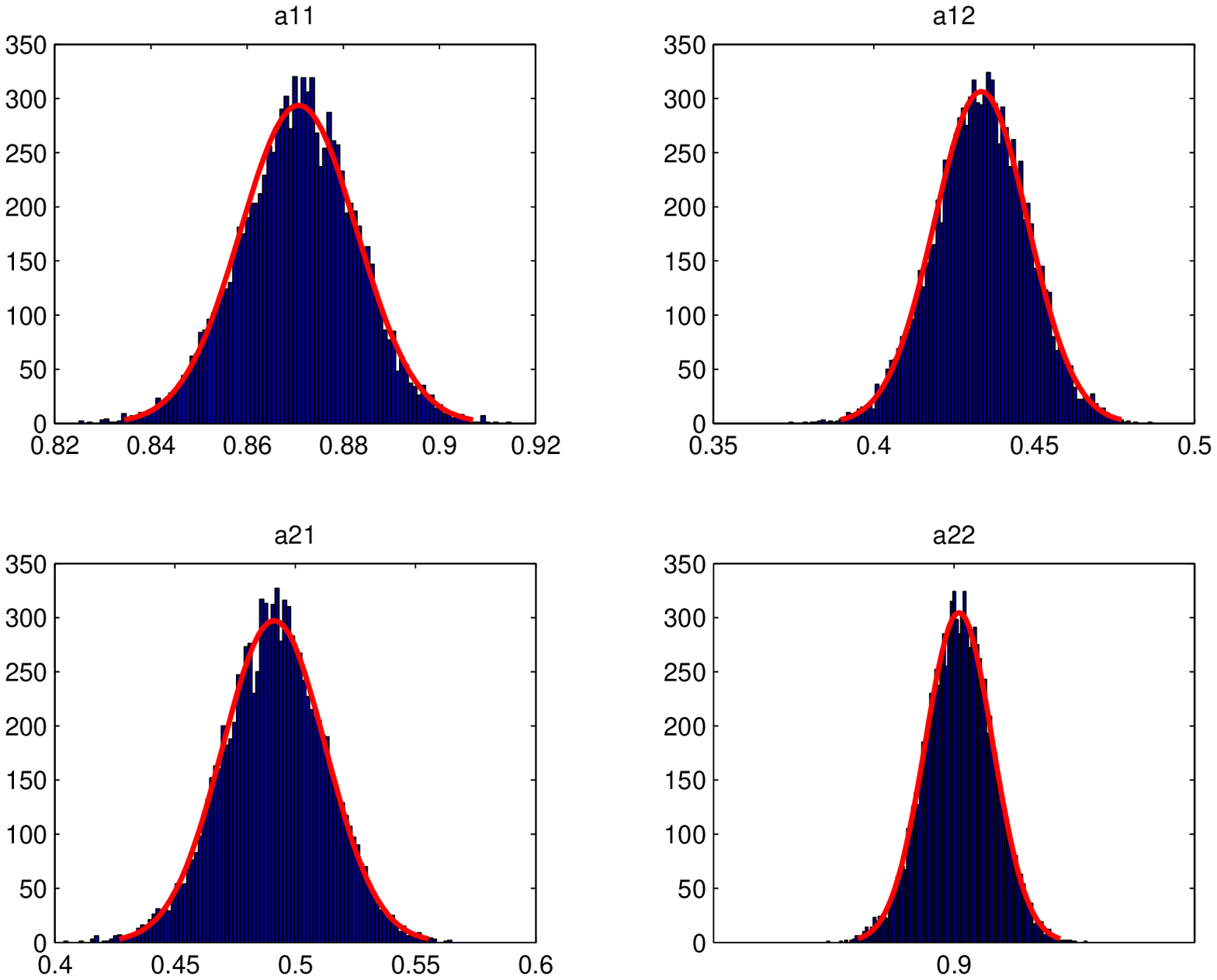}&
\includegraphics[width=70mm,height=55mm]{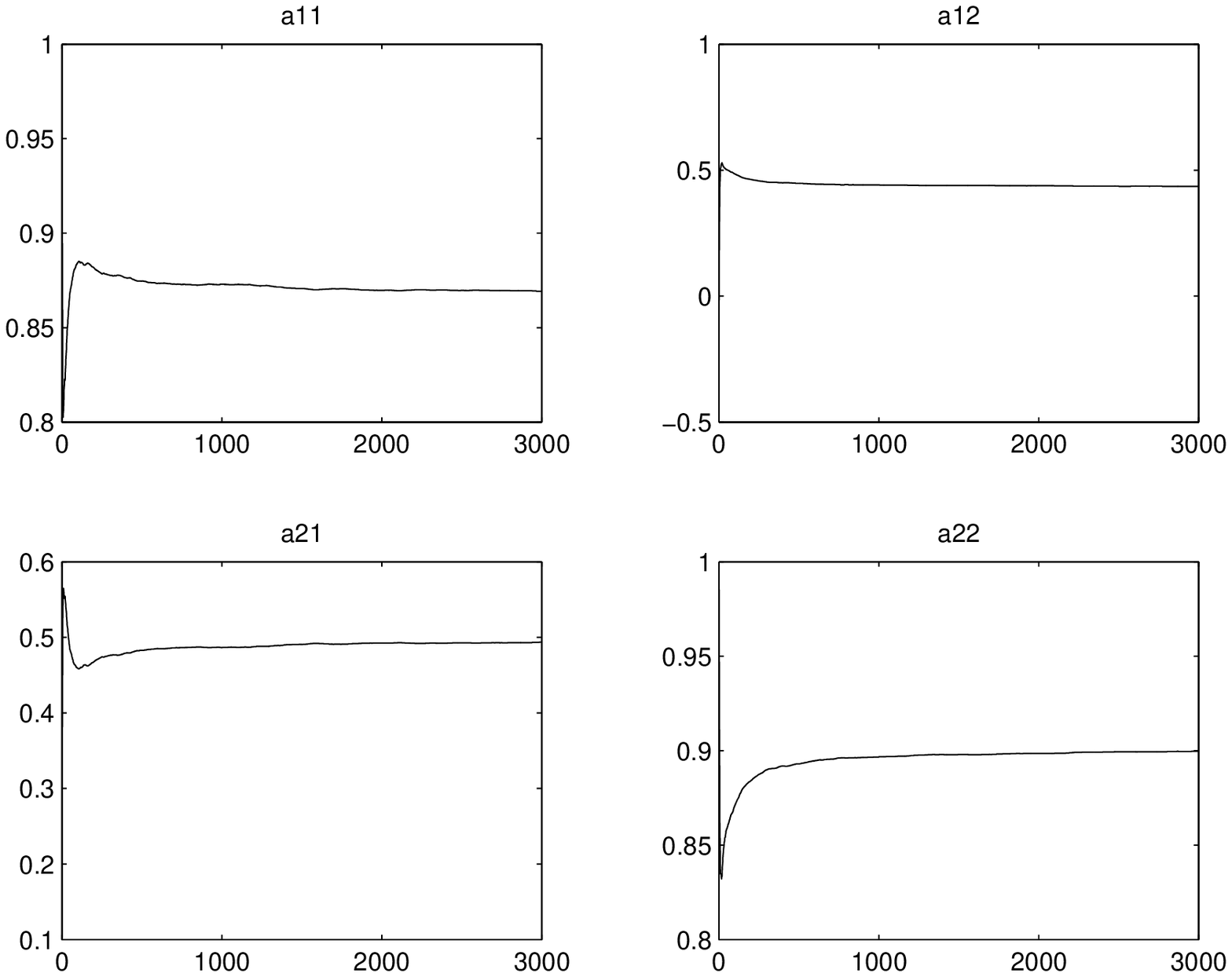} \\
Figure-$2$ The histograms of  the &
Figure-$3$ Convergence of the  empirical  \\
 samples of mixing elements $a_{ij}$& expectations  of $a_{ij}$ after $1000$ iterations
\etabu

\setlength{\tabcolsep}{0.4cm}
\btabu{@{\hspace{0.1mm}}c@{\hspace{0.1mm}}c}
\includegraphics[width=70mm,height=55mm]{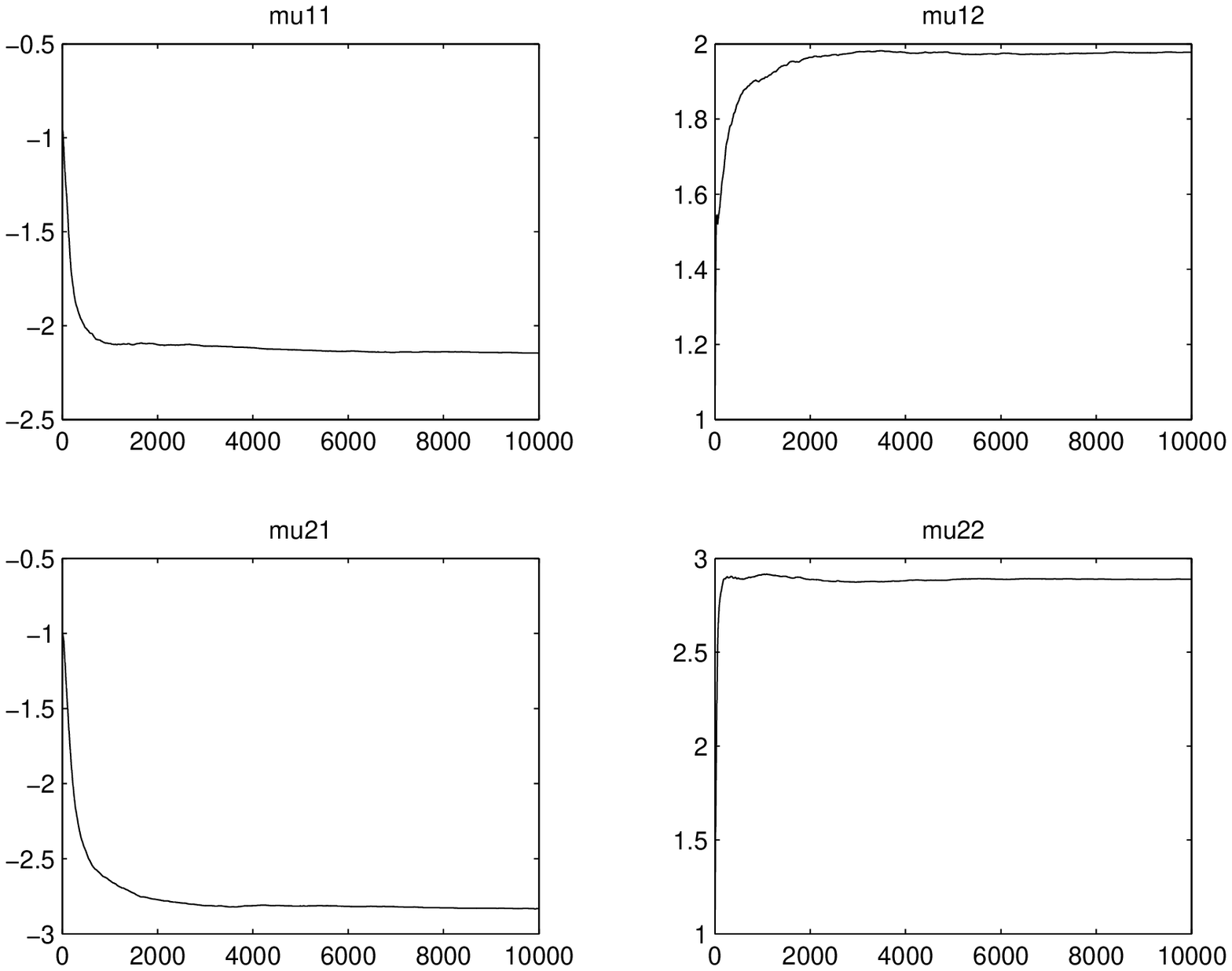}&
\includegraphics[width=70mm,height=55mm]{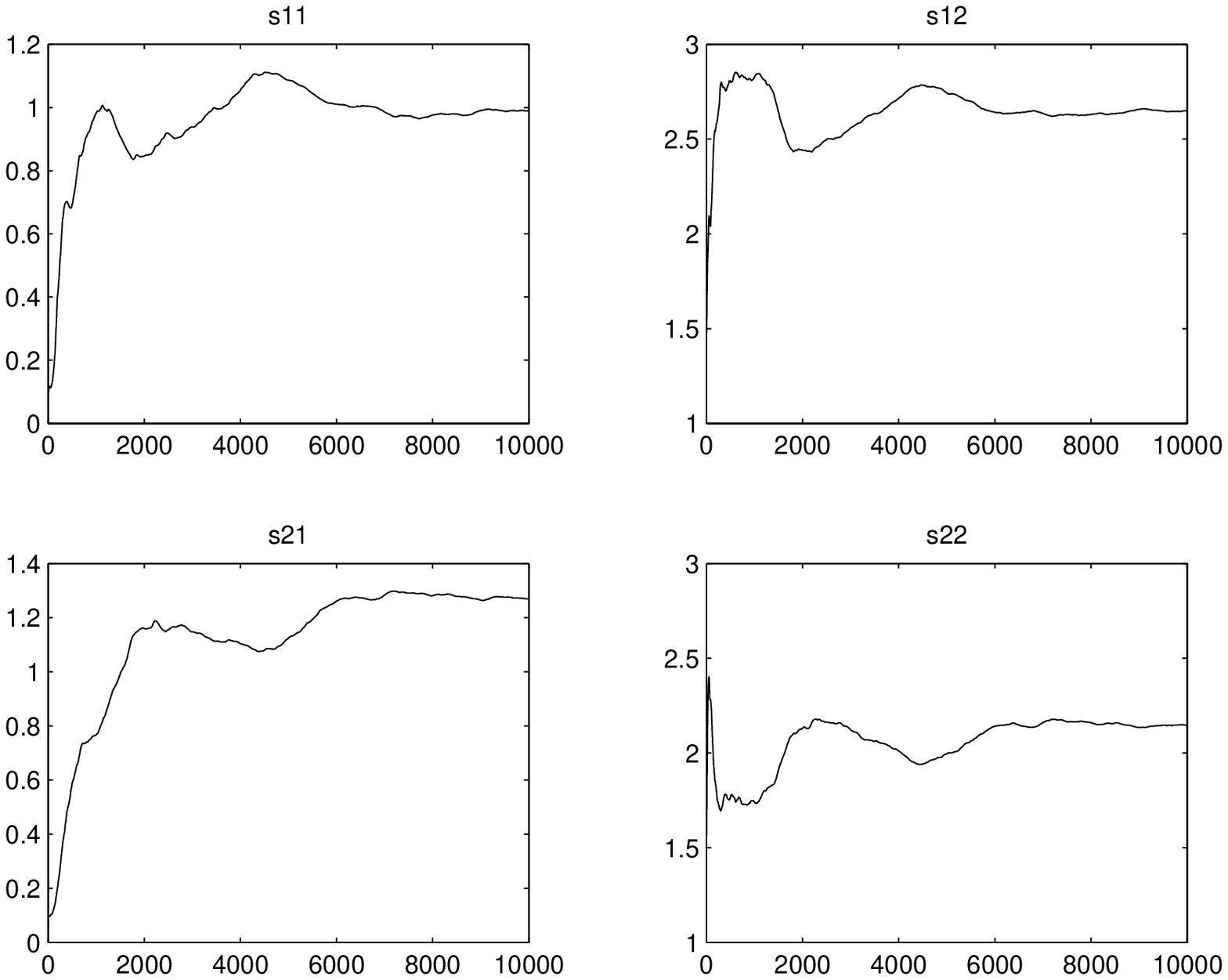} \\
Figure-$4$ Convergence of the empirical  &
Figure-$5$ Convergence of the empirical  \\
  expectations of the means $m_{ij}$ & expectations of the variance $\sigma_{ij}^2$
\etabu

\bcc
\vspace{0.5cm}
\setlength{\tabcolsep}{0.4cm}
\btabu{c}
\includegraphics[width=80mm,height=50mm]{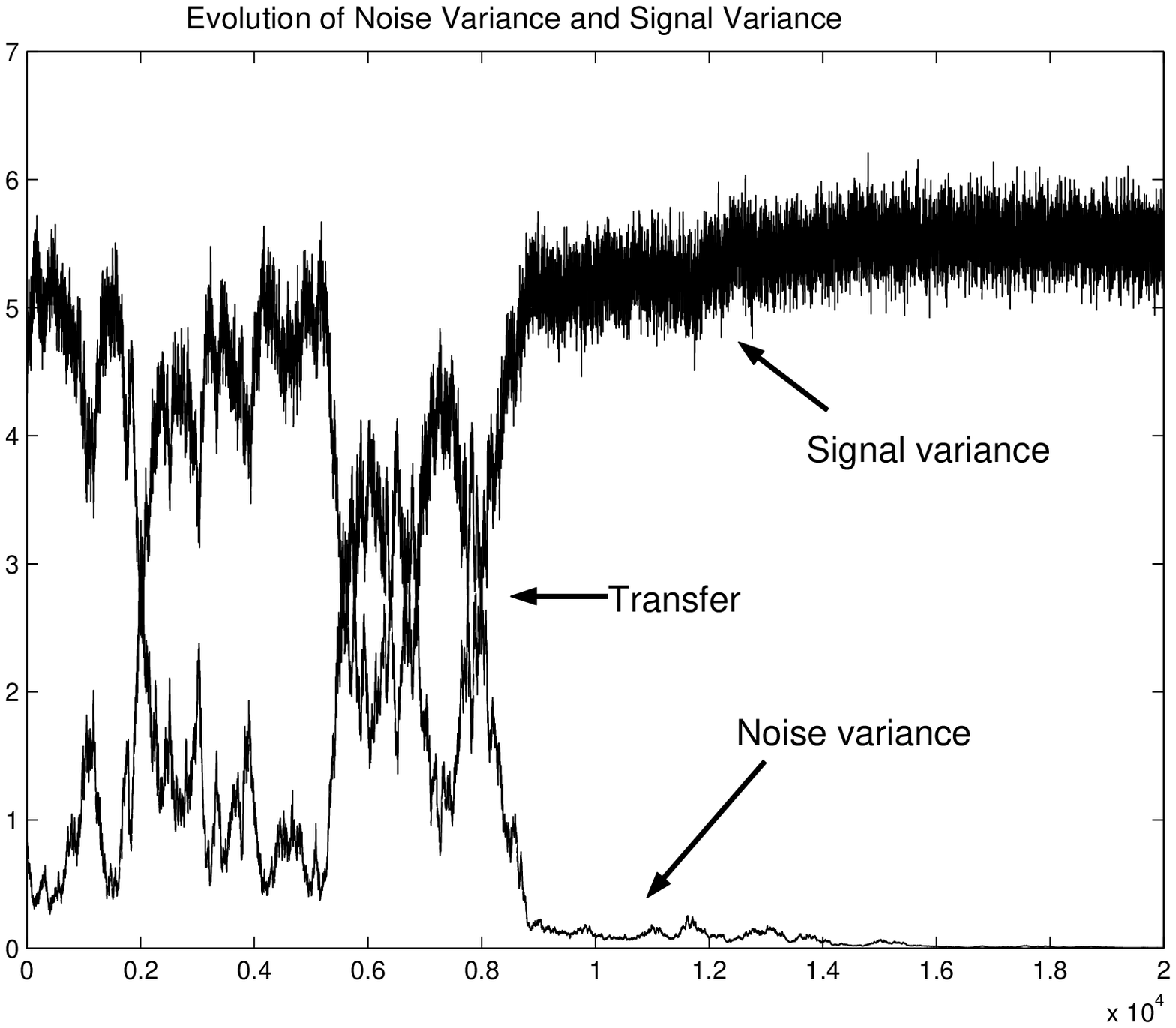} \\
\small{Figure-$6$. The transfer of variances between $\Rbe(1,1)$ }\\
\small{and the variance $\sigma_{11}^2$ after $80000$ iterations} 
\etabu 
\ecc 


\vspace{0.2cm}
We test our algorithm on real data. The first source is a satellite image of an earth region and the second source represents the clouds (First column of figure $7$). The mixed  images are shown in the second column of figure $7$. The results of the algorithm are illustrated in the third column of figure $7$ where the sources are successfully separated. The figure $8$ illustrate the joint segmentation of the sources. We note that the results of the two segmentations are the same as the results which can be obtained if we apply the segmentation on the original sources. 

\vspace{0.5cm}
\setlength{\tabcolsep}{0.4cm}
\btabu{@{\hspace{2mm}}c@{\hspace{2mm}}c@{\hspace{2mm}}c}
\includegraphics[width=30mm,height=30mm]{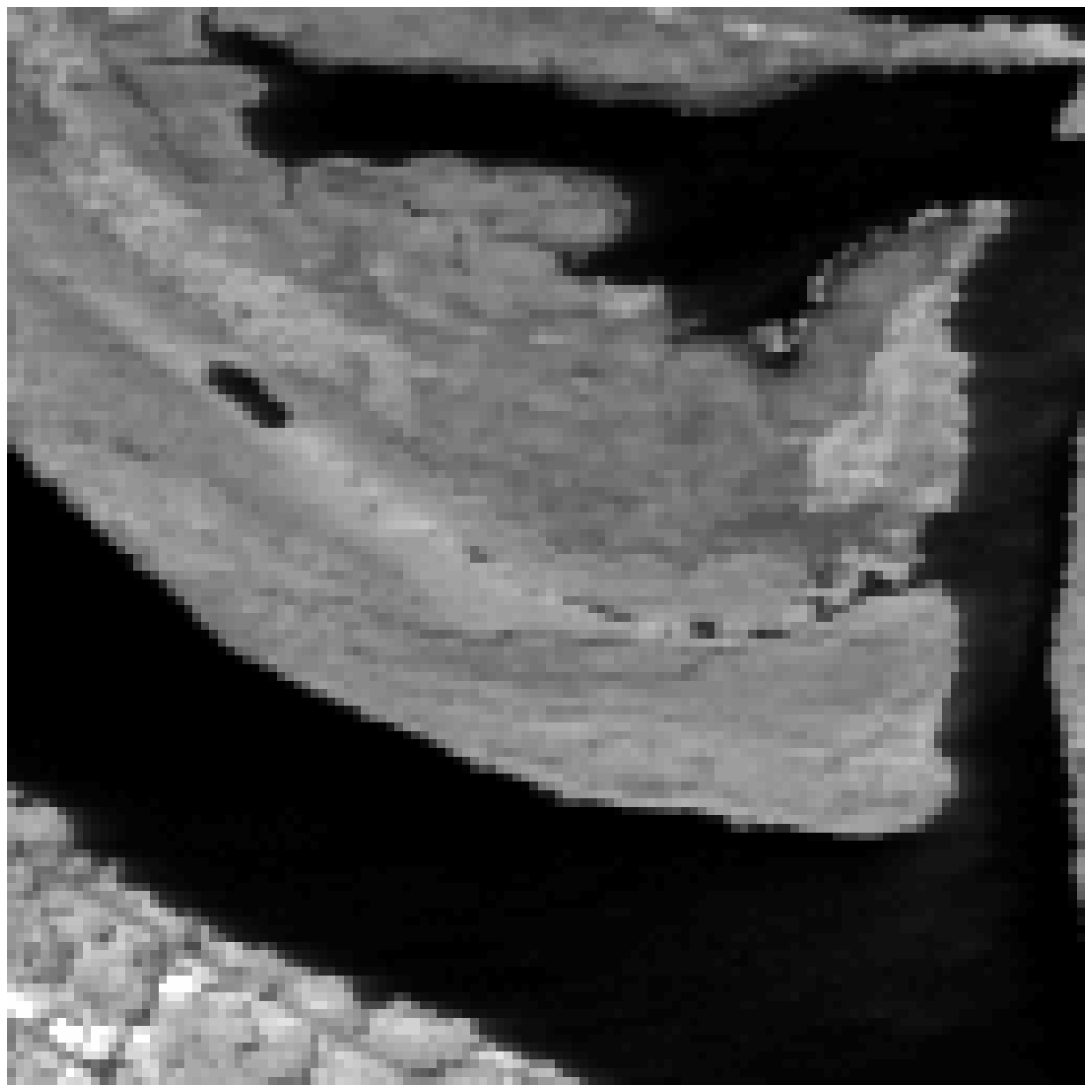} &
\includegraphics[width=30mm,height=30mm]{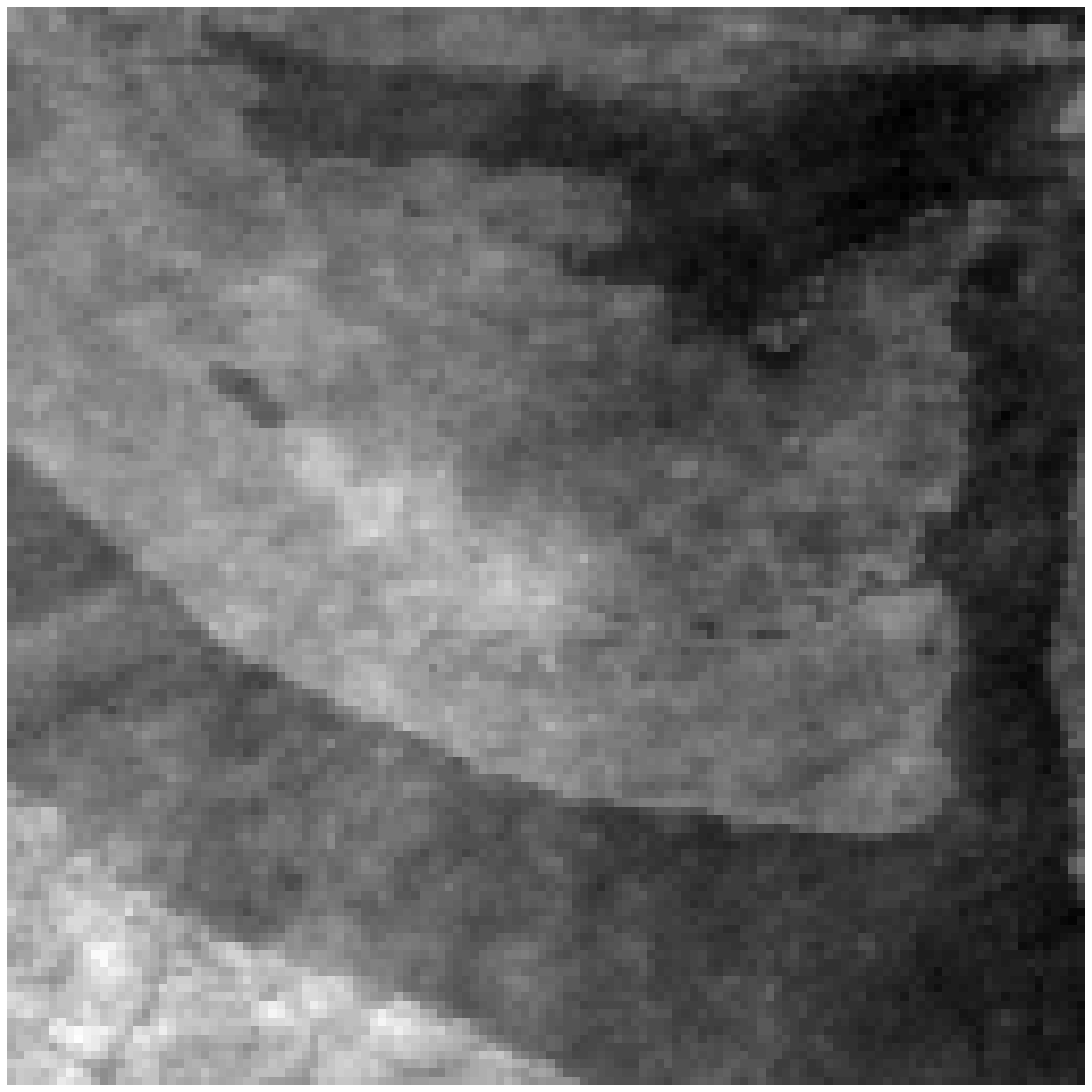} &
\includegraphics[width=30mm,height=30mm]{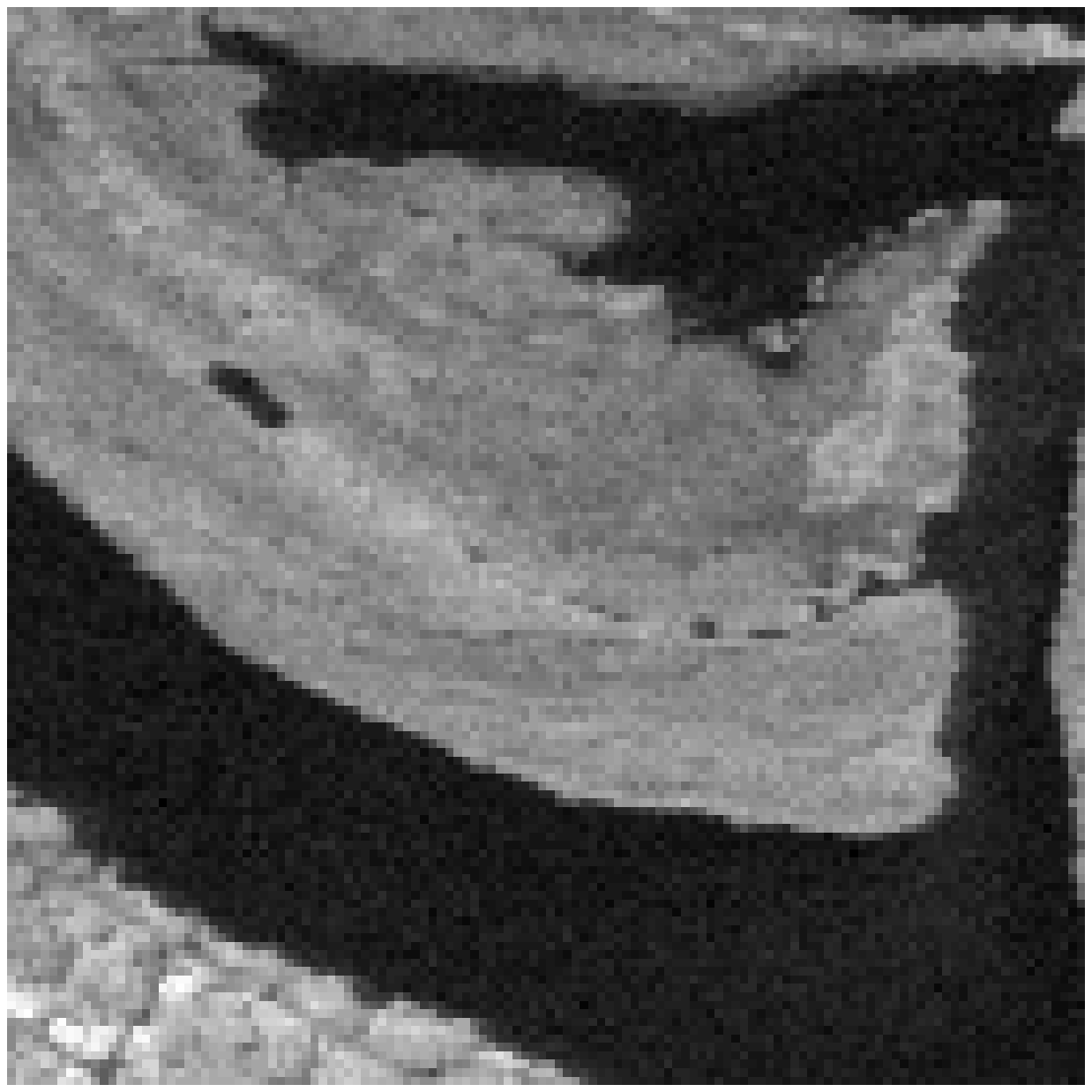} \\
~\\
\includegraphics[width=30mm,height=30mm]{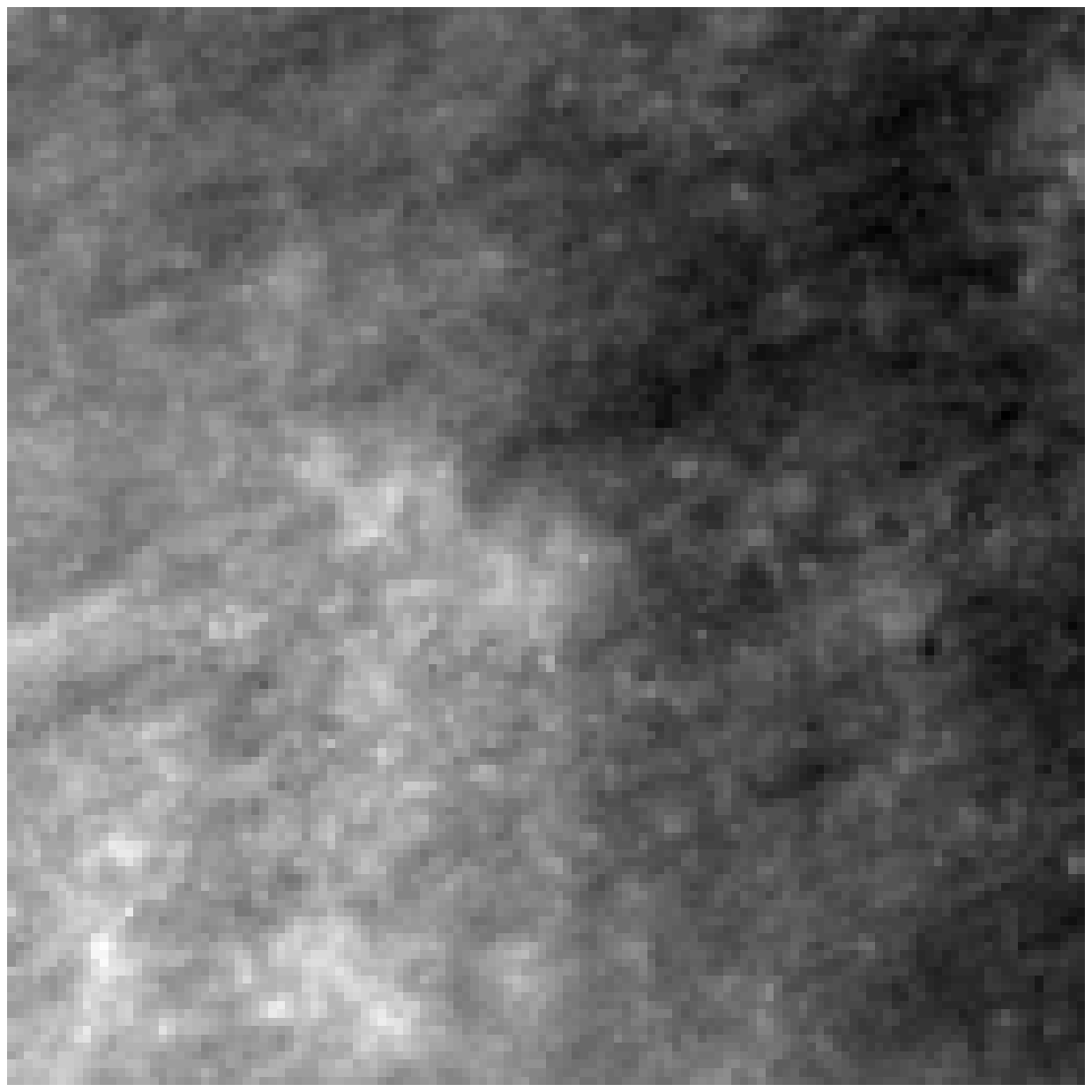} &
\includegraphics[width=30mm,height=30mm]{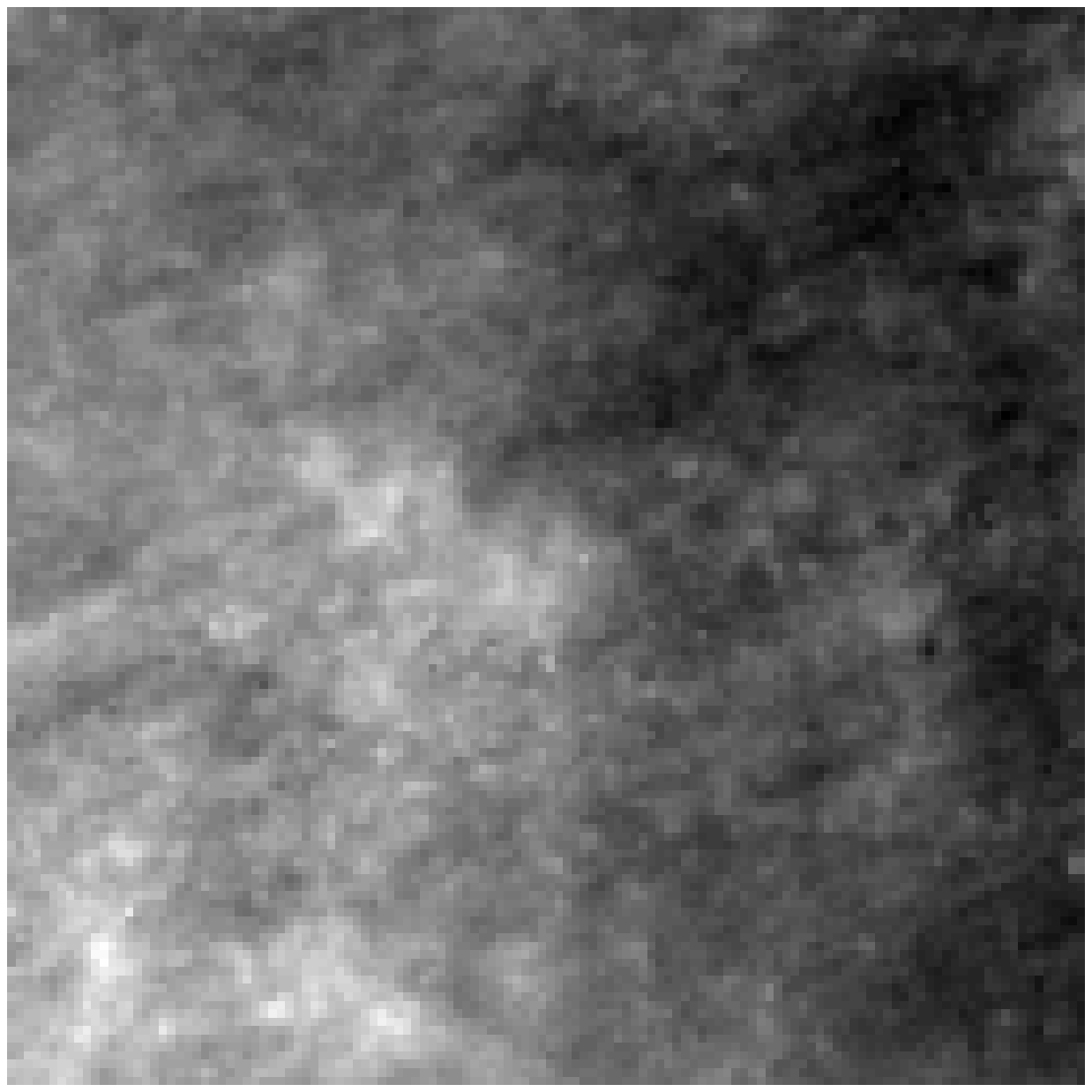} &
\includegraphics[width=30mm,height=30mm]{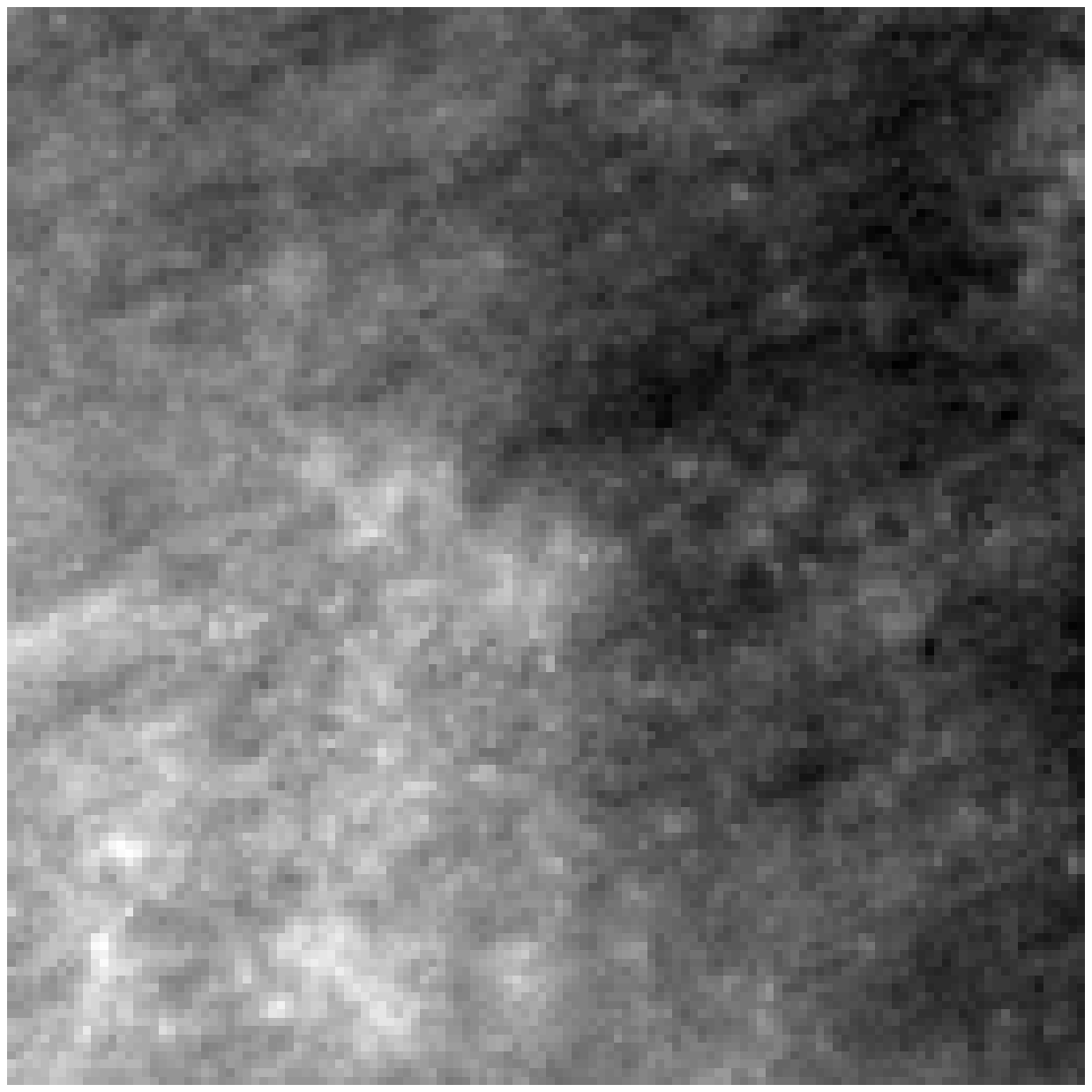}  \\
($a$) & ($b$) & ($c$)
\etabu

\vspace{0.2cm}
\small{Figure.$7$: (a) Original sources, (b) Mixed sources and (c) Estimated sources}

\vspace{0.7cm}
\setlength{\tabcolsep}{0.4cm}
\btabu{@{\hspace{2mm}}c@{\hspace{2mm}}c}
\fbox{\includegraphics[width=40mm,height=30mm]{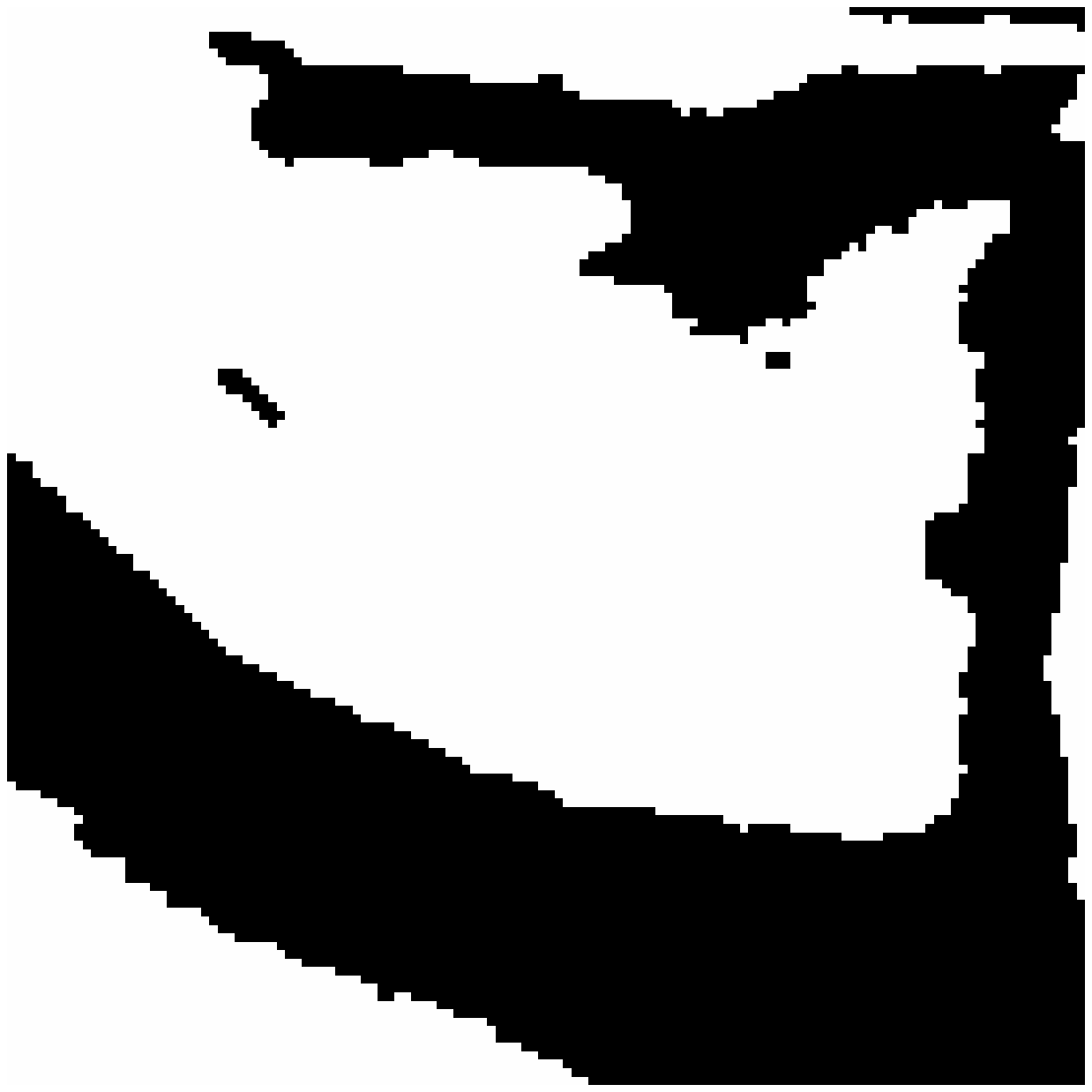}} &
\fbox{\includegraphics[width=40mm,height=30mm]{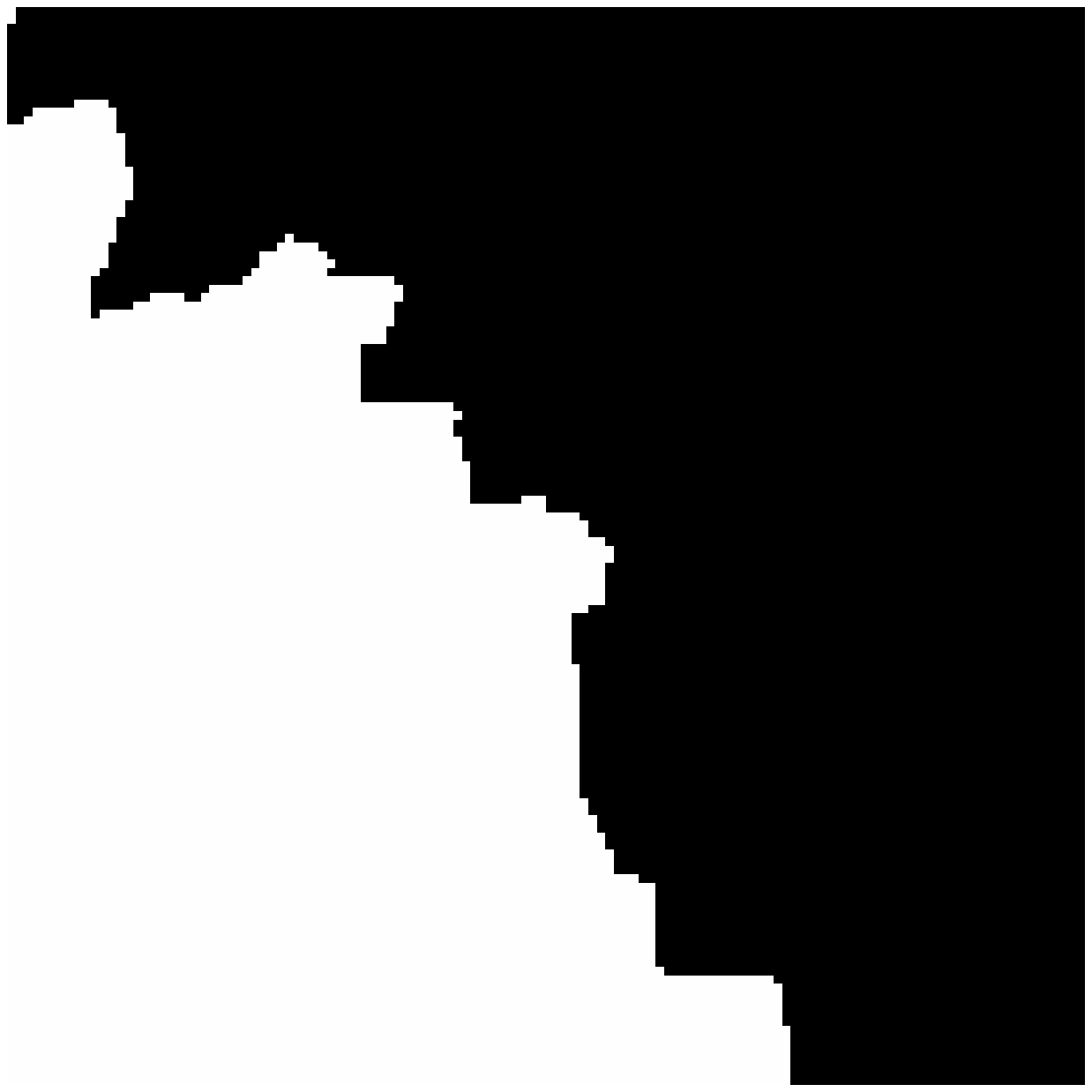}}
\etabu

\begin{center}
\small{Figure $8$: Segmented images}
\end{center}

\section{VI. Conclusion}
In this contribution, we propose an MCMC algorithm to jointly estimate the mixing matrix and the parameters of the hidden Markov fields. The problem  has an interesting natural hidden variable structure leading to  a two-level data augmentation procedure. The observed images are embedded in a wider space composed of the observed images, the original unknown images and  hidden discrete fields modelizing a second attribute of the images and allowing to take into account a Markovian structure. The problems of identifiability and degeneracies are mentioned and discussed. In this work the number of sources and the number of the discrete values of the hidden Markov field are assumed to be known. However, the implementation of the algorithm could be extended to involve the reversible jump procedure on which we are working.

{\small 
\bibliographystyle{ieeeji}
\bibliography{biben,revuedef,revueabr,baseAJ,baseKZ,gpipubli,bss,amd,amd97,ss,cardozo,mackay}
}

\edoc